\documentclass[useAMS,usenatbib]{mn2e}

\usepackage{graphicx}
\usepackage{amsmath}
\usepackage{amsfonts}

\title[Asymptotic, non-linear solutions for ambipolar diffusion in one dimension]{Asymptotic, non-linear solutions for ambipolar diffusion in one dimension}
\author[Jaime H. Hoyos, Andreas Reisenegger, and Juan A. Valdivia]{Jaime H. Hoyos$^{1}$ $^{2}$ $^{3}$ \thanks{E-mail:
jhhoyos@udem.edu.co (JHH); areisene@astro.puc.cl (AR); alejo@macul.ciencias.uchile.cl (JAV)}, Andreas Reisenegger$^{2}$, and Juan A. Valdivia $^{4}$
\footnotemark[0]\\
$^{1}$Departamento de Ciencias B$\acute{a}$sicas, Universidad de Medell$\acute{\i}$n, Cra $87$ N$^o$~$30-65$, Medell$\acute{\i}$n, Colombia\\
$^{2}$Departamento de Astronom$\acute{\i}$a y Astrof$\acute{\i}$sica, Pontificia Universidad Cat$\acute{o}$lica de Chile, Casilla $306$, Santiago $22$, Chile\\
$^{3}$Astrophysikalisches Institut Potsdam, An der Sternwarte $16$, D-$14482$, Potsdam, Germany\\
$^{4}$Departamento de F$\acute{\i}$sica, Facultad de Ciencias, Universidad de Chile, Casilla $653$, Santiago, Chile}
\begin{document}

\date{Accepted 2010 June 22. Received 2010 June 14; in original form 2010 March 21}

\pubyear{2010}

\maketitle

\begin{abstract}
We study the effect of the non-linear process of ambipolar
diffusion (joint transport of magnetic flux and charged particles
relative to neutral particles) on the long-term behavior of a 
non-uniform magnetic field in a one-dimensional geometry.
Our main focus is the dissipation of magnetic energy inside
neutron stars (particularly magnetars), but our results have a
wider application, particularly to the interstellar medium and the
loss of magnetic flux from collapsing molecular cloud cores. Our
system is a weakly ionized plasma in which neutral and
charged particles can be converted into each other through
nuclear beta decays (or ionization-recombination processes).
In the ``weak-coupling'' limit of infrequent inter-particle
interactions, the evolution of the magnetic field is controlled by the beta
decay rate and can be described by a non-linear partial
integro-differential equation. In the opposite,
``strong-coupling'' regime, the evolution is controlled by the
inter-particle collisions and can be modelled through a
non-linear diffusion equation. We show numerically that, in both
regimes, ambipolar diffusion tends to spread out the magnetic
flux, but, contrary to the normal Ohmic diffusion, it
produces sharp magnetic field gradients with associated current
sheets around those regions where the magnetic field is weak.
\end{abstract}


\section{Introduction}

Ambipolar diffusion is the joint drift of charged particles
and the associated magnetic flux with respect to the neutral
particles in a partially ionized plasma. \citet{MestelSpitzer56}
first proposed it in order to explain the loss of magnetic flux
from the dense cores of molecular clouds, required for the
formation of stars, starting an active field of research in this
area. Later, it was suggested to also play a role in the decay of
the magnetic fields of neutron stars
(\citealp{J-87,H-91,P-92,GR-92}, hereafter GR-92) which became
particularly relevant with the identification of ``magnetars'',
neutron stars whose main power source appears to be the
dissipation of their magnetic field \citep{DT-92,TD-96,ACT-04}.

In a previous paper (\citealt{HRV-08}; hereafter Paper I), we
established a multifluid formalism in which it is possible to
study the long-term evolution of magnetic fields in neutron stars
(see \citealt{R-09} for a discussion of the main properties of the
magnetic field equilibria and their subsequent long-term
evolution). In that work, and following the ideas developed by
GR-92, we included the effects of
several physical processes that are also relevant for star
formation and protoplanetary disks, including {\it ambipolar
diffusion}, {\it Hall drift} (non-dissipative advection of the magnetic field
by the associated electrical current), and {\it ohmic diffusion}
(dissipation of currents through the electrical resistivity).

Here we continue this study and concentrate our analysis on the
long-term evolution of the magnetic field caused by 
ambipolar diffusion aided by beta decays. Following the same
philosophy of Paper I, and as a first approach to the
understanding of our general formalism, we focus on a simplified,
one-dimensional configuration in which the magnetic field points
in one Cartesian direction $z$ but varies only along an orthogonal
direction $x$, i.~e., $\vec B=B(x,t)\hat z$. Such models have
also been considered in several studies of ambipolar diffusion in
the interstellar medium
\citep{MouschoviasPaleologou81,Shu-83,BZ-94}, although some of the
assumptions differ from case to case. In our analysis, we
consider separately two relevant limits, similar in spirit,
though not exactly equivalent to those of
\citet{MouschoviasPaleologou81}:
\begin{itemize}
\item In the {\it weak coupling} limit, there are few collisions
between the particles, and the beta decays proceed slowly.
Therefore, the particles can reach the diffusive equilibrium
easily, but it takes much longer to reach the chemical
equilibrium. \item In the {\it strong coupling} limit, there are
many collisions between the particles, and the beta decays proceed
fast, so the (local) chemical equilibrium is reached much
more quickly than the diffusive equilibrium.
\end{itemize}

For each of these cases, we find that the long-term evolution
of the magnetic field can be modelled by a single equation that
gives the time-derivative $\partial B/\partial t$ at a given
instant $t$ only in terms of the configuration of the magnetic
field at the same instant, $B(x,t)$. This makes it easy to carry
out numerical simulations of the evolution of some selected
non-linear magnetic field profiles, and even find some exact, analytical solutions.

In Sect.~\ref{1dmodel} of this paper, we briefly review the
one-dimensional model of neutron star magnetic field
evolution introduced in Paper I, paying particular attention
to its characteristic evolutionary time-scales.

In Sects.~\ref{asy1} and \ref{asy2}, we obtain the equations for the long-term, asymptotic magnetic field evolution
promoted by ambipolar diffusion in each of the two opposite
regimes mentioned above, and we make numerical simulations of the
evolution of different initial magnetic field configurations. 
We show that, in both cases, the magnetic flux of a given sign
tends to spread out, but singularities develop at the null points
where regions with different signs meet, as previously found by
\citet{BZ-94}. In the weak coupling case, these singularities
correspond to current sheets that are dissipated by resistive
effects, in this way leading to reconnection. In the strong
coupling case, the singularities have a somewhat different
character (a smoothly diverging current density) and 
might lead directly to reconnection even in the case of no ohmic resistivity (but see \citealp{H-03a,H-03b}).
Finally, in Sect.~\ref{conc}, we give the main conclusions of our study.

\section{One-dimensional model and evolutionary time-scales in neutron stars}
\label{1dmodel}

We model the neutron star interior as an electrically neutral and
slightly ionized plasma composed of three mobile, strongly
degenerate, particle species: neutrons $(n)$, protons $(p)$, and
electrons $(e)$. We account for binary collisions and weak
interactions (causing nuclear beta decays) between the particles,
and allow for strong interactions between neutrons and
protons by writing each of their chemical potentials as a function
of both of their number densities: $\mu_{n,p}=\mu_{n,p}(n_n,n_p)$, while considering the electrons as an ideal, relativistic
Fermi gas, whose chemical potential is a function only of their
number density, $\mu_e=\mu_e(n_e)$.

We study a one-dimensional geometry in which the magnetic field
points in one Cartesian direction $z$, but varies only along an
orthogonal direction $x$ as $\vec{B}(\vec{r},t)=B(x,t)\hat{z}$,
and assume that all physical variables vary only along $x$. Since,
in neutron star conditions, the ratio of the magnetic
pressure $B^2/8\pi$  to the pressure of the charged particles
is very small, the magnetic force causes only small perturbations
to the hydrostatic equilibrium state of the non-magnetized star.

These assumptions are generally not true in molecular cloud
cores, where the ionization fraction tends to be extremely low,
whereas the magnetic field can be near equipartition with the
neutral gas pressure. In this sense, our derivation will be valid
only for the case of neutron stars, although we will see that some
of the results agree with those of other authors, obtained under
somewhat different assumptions. A more general treatment appears
to be difficult and not to yield simple results.

For the reasons stated, we consider a non-magnetized, fixed
background system in hydrostatic and chemical equilibrium and
introduce small perturbations to the number density of each
species $i$ as $n_i(x,t)= n_{0i}(x)+\delta n_i(x,t)$, with the
subscript zero labelling the background number densities and
$|\delta n_i(x,t)|\ll n_{0i}(x)$. The associated chemical
potential perturbations are given by $\mu_{i}=\mu_{0i}+\delta
\mu_{i}$.

The long-term magnetic field evolution implies small particle
velocities that change over long time-scales, much longer than the
very short dynamical times that are only relevant shortly after
the formation of the star (i.e, sound or Alfv\'en time-scales).
Therefore, at all times we use a slow-motion approximation in
which we neglect the acceleration terms in the equations of motion
for the particles.\footnote {This approximation should be
roughly true also for star formation, and was indeed assumed by
some authors \citep{Shu-83,BZ-94}, but not by
\citet{MouschoviasPaleologou81}, who dropped the charged-particle
pressure, but kept the inertial terms, and therefore obtained a
differential equation of second rather than first order in time
for the magnetic field evolution.}

Taking account of all these considerations, the system of
non-linear partial differential equations governing the evolution
is (see Paper I for the derivation)
\begin{equation}
\label{bz}
\frac{\partial B}{\partial t}=-\frac{\partial}{\partial x}\left(v_{c}B-\frac{c^2}{4\pi\sigma_0}\frac{\partial B}{\partial x}\right),
\end{equation}
\begin{equation}
\label{nB}
\frac{\partial \delta n_B}{\partial t}= -\frac{\partial}{\partial x}\left(n_{0n}v_n+n_{0c}v_{c}\right),
\end{equation}
\begin{equation}
\label{nc}
\frac{\partial \delta n_c}{\partial t}=-\frac{\partial}{\partial x}\left(n_{0c}v_{c}\right)-\lambda \left(\delta \mu_c - \delta \mu_n\right),
\end{equation}
where
\begin{eqnarray}
\label{vn}
v_n=-\frac{1}{\alpha n_{0n}}\left[n_{0n}\mu_{0n}\frac{\partial}{\partial x}\left(\frac{\delta \mu_n}{\mu_{0n}}\right)
+n_{0c}\mu_{0c}\frac{\partial}{\partial x}\left(\frac{\delta \mu_c}{\mu_{0c}}\right)\right.\nonumber\\+\left.\frac{\partial}{\partial x}\left(\frac{B^2}{8\pi}\right)\right],
\end{eqnarray}
and
\begin{equation} \label{vA}
v_A=-\frac{1}{n_{0n}n_{0c}\gamma_{cn}}\left[n_{0c}\mu_{0c}\frac{\partial}{\partial
x}\left(\frac{\delta
\mu_c}{\mu_{0c}}\right)+\frac{\partial}{\partial
x}\left(\frac{B^2}{8\pi}\right)\right].
\end{equation}

The magnetic field evolution is given by Eq.~(\ref{bz}), where
$v_c$ denotes the velocity of the charged particles (the same for
electrons and protons), $v_n$ is the velocity of the neutrons, and
$v_A\equiv v_c-v_n$ is the ambipolar diffusion
velocity.\footnote{Strictly speaking, all these are the
$x$-components of the respective velocities, which cause changes
in the distributions of particles and magnetic flux. The
charged-particle velocities must also have a $y$-component,
responsible for the current that acts as the source of the
magnetic field.} The first term on the right-hand side of this
equation causes an advection of the magnetic flux with a velocity
$v_c$, while the second term describes the ohmic diffusion of the
field, where $\sigma_0$ is the electrical conductivity. In neutron
star core conditions, the electrical conductivity is $\sigma_0
\sim 10^{28}s^{-1}$, thus, the
evolution of the large-scale magnetic field through
ohmic diffusion proceeds very slowly, with a time-scale
$t_{ohmic}\sim 10^{11}~ \mathrm{yr}$, longer than the age of the
universe \citep{Bay-69}. Qualitatively similar conditions
hold in essentially all astrophysical settings. Therefore, in the
rest of this paper, we neglect the ohmic term in the magnetic
field evolution equation and focus only on the magnetic evolution
due to the advective term. However, the time-scale for ohmic
diffusion scales with the square of the characteristic length of
the magnetic field variations (see Paper I), thus, the influence
of ohmic dissipation can be important in regions with strong
spatial magnetic field variations.

The evolution of the particle number density perturbations, for
the different species, is given by Eqs.~(\ref{nB}) and (\ref{nc}),
where $\delta n_e=\delta n_p \equiv \delta n_c$ is the
perturbation of the charged particle number density keeping
charge neutrality and $\delta n_B=\delta n_n+\delta n_c$ is the
perturbation of the baryon number density. Weak interactions cause
beta decays (conversion of charged particles into neutrons and the
opposite) that tend to reduce deviations from the {\it chemical
equilibrium} state between charged particles and neutrons, with a
net rate coefficient $\lambda$ and $\delta \mu_c\equiv \delta
\mu_e+\delta \mu_p$. The chemical equilibrium is achieved when
$\delta \mu_c=\delta \mu_n$.

Continuing our description, Eq.~(\ref{vA}) shows that the
ambipolar diffusion velocity $v_A$ is controlled by the collision
rate between the charged particles and neutrons, which is
proportional to the parametre $\gamma_{cn}$. This velocity is
driven by the Lorentz force but choked by the charged particle
pressure gradient it produces \citep{P-92,GR-92}.

Regarding the neutrons, we see from the right-hand side of
Eq.~(\ref{vn}) that they move with a velocity $v_n$ as long
as 
the system is not in a state of {\it magneto-hydrostatic
equilibrium}. In this state, the Lorentz force is balanced by the
pressure gradients of all the particles (charged particles and
neutrons). We also see that the neutron velocity is controlled by
the parametre $\alpha$, whose meaning we explain in the next
paragraph.

Since we are interested in a numerical solution of our equations,
we have to take into account that modelling the true dynamical
time-scales (sound or Alfv\'en time-scales of milliseconds to
seconds) would require a time step many orders of magnitude
shorter than that required to simulate the long-term evolution,
making the simulation computationally unfeasible. For this
reason, in Paper I we introduced a {\it slow-motion
approximation}, neglecting the acceleration terms and instead
introducing a small, artificial, friction-like force acting on the
neutrons (the most abundant species) of the form $-n_{0n}\alpha
v_n$, where the parametre $\alpha$ is chosen in such a way that
the time to reach magneto-hydrostatic equilibrium is long enough
for the numerical code to be able to deal with it (and therefore,
much longer than the real dynamical times), but shorter than the
long time-scales over which the magnetic field evolves. We showed
in Paper I that the latter time-scales, which are the
astrophysically interesting ones, are unaffected by the choice of
$\alpha$.

We require the conservation of both the magnetic flux
$\Phi=\int_0^{d}B~dx$ and the baryon number perturbation $\delta
N_B=\int_0^{d}\delta n_B~dx$ during the evolution of our
system, which spans the segment $0\leq x\leq d$. We ensure this
through the boundary conditions

\begin{equation}
v_{c}(x=0,t)=v_{c}(x=d,t)=0,
\end{equation}

\begin{equation}
v_n(x=0,t)=v_n(x=d,t)=0,
\end{equation}

\begin{equation}
\label{boundaries}
\frac{\partial B}{\partial x}(x=0,t)=\frac{\partial B}{\partial x}(x=d,t)=0.
\end{equation}

The system of Eqs.~(\ref{bz}-\ref{nc}) describes the evolution of
three coupled variables: the magnetic field, the charged particle
density perturbation, and the baryon density perturbation. In
Paper I, we estimated the three associated characteristic
evolutionary time-scales of this set of equations 
corresponding to exponentially decaying eigenmodes in the linear
approximation and showed that they characterize the approach
to three successive equilibrium states. Here we briefly describe
the evolutionary stages and summarize the relevant time-scales
with the main goal of establishing the basic ideas that will be
used in our subsequent analysis (see Paper I for more details
about their derivation).

The shortest time corresponds to the approach of the
magneto-hydrostatic equilibrium, controlled by $\alpha$ in our
model, as already explained.

In a longer time, two alternative processes compete:
\begin{enumerate}
\item The particle species move relative to each other, controlled
by the inter-particle collisions, in a tendency to reach a
diffusive equilibrium state in which the fluid forces acting on
each species separately are balanced. \item The weak interactions
convert particles from one species into another, tending towards a
chemical equilibrium state.
\end{enumerate}

If the relative motion of charged and neutral particles
proceeds much faster than the conversion from one into another, a
diffusive equilibrium is achieved in the system, characterized by
the balance equations
\begin{equation}
\label{charg-diff}
0=n_{0c}\mu_{0c}\frac{\partial}{\partial x}\left(\frac{\delta \mu_c}{\mu_{0c}}\right)
+\frac{\partial}{\partial x}\left(\frac{B^2}{8\pi}\right),
\end{equation}
and
\begin{equation}
\label{neut-diff}
0=n_{0n}\mu_{0n}\frac{\partial}{\partial x}\left(\frac{\delta \mu_n}{\mu_{0n}}\right).
\end{equation}
The inter-particle collision frequency controls the time-scale to
reach the diffusive equilibrium, which is given by Paper I,
\begin{equation}
\label{tnccgen}
t_{drag}\sim \frac{\gamma_{cn}L^2}
{\left(\frac{\partial \mu_{0n}}{\partial n_{0B}}\right)_{n_{0c}}}
\sim~4.5~\times 10^{-1}~ L_5^2~T_8^2 ~ \mathrm{yr}.
\end{equation}
where $L_5\equiv L/(10^5 \mathrm{cm})$ and $T_8\equiv T/(10^8 \mathrm{K})$. If the beta decays are much faster, the chemical equilibrium state $\delta \mu_n=\delta \mu_c$ is
reached in a time-scale controlled by the beta decay rate,
\begin{equation}
\label{dmu1gen}
t_{weak}\sim \frac{n_{0c}}{\lambda n_{0n} \left(\frac{\partial \mu_{0n}}{\partial n_{0B}}\right)_{n_{0c}}}\sim 4.3 ~ \times 10^5~ T_8^{-6} ~\mathrm{yr}.
\end{equation}

Thus, we have two relevant limits:
\begin{enumerate}
\item In the {\it weak coupling limit}, $t_{drag} \ll t_{weak}$
and the weak interactions operate slowly, therefore, the system
relaxes to a diffusive equilibrium in a time-scale $t_{drag}$, but
remains out of chemical equilibrium. \item In the opposite, {\it
strong coupling limit}, $t_{drag} \gg t_{weak}$, the
inter-particle collisions are very frequent which impedes a fast
achievement of the diffusive equilibrium, while the system can
relax to the chemical equilibrium in the time-scale $t_{weak}$.
\end{enumerate}
The transition between these limits is achieved when $t_{drag}\sim t_{weak}$. From Eqs.~(\ref{tnccgen}) and (\ref{dmu1gen}) we infer that the condition for this transition is $T_8 \sim 5.6~L_5^{-1/4}$, which gives $t_{drag}\sim t_{weak}\sim 14~ L_5^{3/2}~ \mathrm{yr}$. 

Since the density perturbations are assumed to be small,
these will not involve large motions of the particles, and
therefore not cause a substantial change in the magnetic flux
distribution. Note, however, that these two equilibria are
incompatible with each other as long as a spatially non-uniform
magnetic field is present, therefore a full equilibrium will only
be reached in a much longer time-scale, on which the magnetic
field is made uniform (in our model) or expelled from the system
(likely more realistic in a true astrophysical setting).
For the determination of this much longer time-scale, on which the
magnetic field does evolve substantially, we again consider the
two opposite regimes discussed above. In the {\it weak coupling
limit}, the system reached the diffusive equilibrium but not the
chemical equilibrium, during the previous stage. During this much
longer stage, the weak interactions slowly convert charged
particles into neutrons in a tendency to reduce the
charged-particle pressure gradient that counterbalances the
magnetic pressure gradient. This causes a slight deviation from
the diffusive equilibrium, producing a joint transport of the
charged particles and the magnetic flux at a small ambipolar
diffusion velocity $v_A$, always keeping the system very close to
diffusive equilibrium. This interplay continues until both the
pressure and magnetic field gradients dissapear, which occurs in a
long time-scale that depends on the magnetic field strength and
the weak interaction rate:

\begin{equation}
\label{tbp2}
t_{ambip}^{(weak)}\sim \beta ~ t_{weak}\sim \frac{8\pi n_{0c}^ 2}{\lambda B^2}
\sim~ 1.7\times~ 10^{9}~ B_{15}^{-2}~ T_8^{-6}~\mathrm{yr},
\end{equation}
with $\beta \equiv 8\pi n_{0c} n_{0n} \left(\partial
\mu_{0n}/\partial n_{0B}\right)_{n_{0c}}/B^2\gg 1$, which is
roughly the ratio of the charged particle pressure to the magnetic
pressure and we defined $B_{15}\equiv B/(10^{15} \mathrm{G})$. 

In the {\it strong coupling limit}, the relative motion of the
charged particles and neutrons is strongly suppressed by the
inter-particle collisions, which delay the diffusive equilibrium
state, while in comparison the chemical equilibrium is reached
quickly. The deviation from diffusive equilibrium promotes, as
before, a joint motion of the charged particles and the magnetic
flux, with a very small ambipolar diffusion velocity $v_A$. This
movement yields the diffusive equilibrium in a very lon
time-scale controlled by the collision rate between particles and
by the magnetic-field amplitude,
\begin{eqnarray}
\label{tambi22p2}
t_{ambip}^{(drag)}\sim \beta ~ t_{drag}\sim \frac{8\pi \left(1+\frac{n_{0c}}{n_{0n}}\right)n_{0B}n_{0c}\gamma_{cn} L^2}{B^2}\nonumber\\~\sim 1.8\times 10^{3}~ B_{15}^{-2}~ L_{5}^2 ~ T_8^{2}~\mathrm{yr}.
\end{eqnarray}

Note that at the transition between the strong and weak coupling limits ($T_8 \sim 5.6~L_5^{-1/4}$), the ambipolar diffusion timescales are of the same order: $t_{ambip}^{(drag)}\sim t_{ambip}^{(weak)} \sim 5.5 \times 10^{4}~ B_{15}^{-2}~ L_{5}^{3/2}~\mathrm{yr}.$ This value corresponds to the shortest possible ambipolar diffusion time, since the relevant timescales increase both towards higher and lower temperatures. Although the Hall effect is not present in our one-dimensional calculations, it is important to asses its potential importance relative to ambipolar diffusion. Consider its timescale,:

\begin{equation}
\label{thall}
t_{Hall} \sim \frac{4\pi n_{0e} L^2}{cB}~\sim 3\times 10^{5}~ B_{15}^{-1}~ L_{5}^2 ~\mathrm{yr},
\end{equation}

compared to the minimum ambipolar diffusion timescale estimated above. We find that, if $B_{15} < 0.2~ L_5^{-1/2}$, the Hall drift is likely to play a dominant role. It might not be important in magnetars, where $B_{15} > 1$, except possibly in reconnection layers, where $L_5\ll 1$. 

In the present paper, we are interested in the details of the
evolution of the magnetic field in the strong and weak coupling limits. Therefore, in the next sections we will obtain the
differential equations modelling the evolution of the magnetic
field in the time-scales given by Eqs.~(\ref{tbp2}) and
(\ref{tambi22p2}).

\section{Magnetic field evolution in the weak coupling limit}
\label{asy1}

\subsection{Derivation of the asymptotic evolutionary equation}

In this section, we derive a single integro-differential
equation that models the magnetic field evolution in the {\it weak
coupling limit} ($t_{drag} \ll t_{weak}$), in which neutral
and charged particles drift easily with respect to each other, and
the main bottleneck is the (slow) rate at which they can be
converted into each other, in order to decrease the
charged-particle pressure gradients that balance the Lorentz
force, impeding the magnetic flux to spread. This limit is not
likely to be relevant in the interstellar medium \citep{Shu-83},
although it roughly corresponds to one of the limits considered by
\citet{MouschoviasPaleologou81}. In neutron stars, it becomes
important at later stages, once their temperature is low enough.

Hereafter, we take the background properties to be homogeneous,
i.e. the variables with a subscript $0$ do not depend on the
position. We assume that the system has already reached the
diffusive equilibrium, which also implies the magneto-hydrostatic
equilibrium. However, since weak interactions are slow, the system
may be out of chemical equilibrium, namely, $\delta \mu_n \neq
\delta \mu_c$. From Eq.~(\ref{neut-diff}) we infer a spatially
uniform distribution of the neutrons,
\begin{equation}
\label{diffusive-eq-neu}
\delta \mu_n=g(t),
\end{equation}
with $g(t)$ an arbitrary function that depends only on the time
variable. From Eq.~(\ref{charg-diff}), the diffusive equilibrium
of the charged particles implies
\begin{equation}
\label{diffusive-eq-cha}
n_{0c}\delta \mu_c+\frac{B^2}{8\pi}=f(t).
\end{equation}

Since $|\delta n_c| \ll n_{0c}$ and the time-scale for the
evolution is $t_{ambip}^{(weak)}\sim L/v_c$, we can compare the
terms in Eq.~(\ref{nc}),
\begin{equation}
\label{orderan} 
\left|\frac{\partial n_c}{\partial t}\right|\sim\frac{\delta n_c}{t_{ambip}^{(weak)}} \ll \frac{n_{0c}v_c}{L}\sim\left|\frac{\partial}{\partial x}(n_{0c}v_c)\right|,
\end{equation}
so we can neglect the time derivative and write this equation as
\begin{equation}
\label{vcasy} 
\frac{\partial v_c}{\partial x}=-\frac{\lambda}{n_{0c}} \left(\delta \mu_c -\delta \mu_n \right).
\end{equation}
Integrating this equation between $x=0$ and an arbitrary internal
point with coordinate $x$, we obtain
\begin{equation}
\label{vcasyint1}
v_c(x,t)=-\frac{\lambda}{n_{0c}}\int_{0}^x\left[\delta \mu_c(x^\prime,t)
-\delta \mu_n(x^\prime,t)\right]dx^\prime,
\end{equation}
where we used the boundary condition $v_c(0,t)=0$. Replacing Eqs.~(\ref{diffusive-eq-neu}) and (\ref{diffusive-eq-cha}) in Eq.~(\ref{vcasyint1}) and using the boundary condition $v_c(d,t)=0$ to eliminate $f(t)$ and $g(t)$, we obtain
\begin{equation}
\label{ambvelr1}
v_c(x,t)=-\frac{\lambda}{dn_{0c}^2}x(d-x)
\left(\frac{1}{d-x}\int_x^d\frac{B^2}{8\pi}dx^\prime
-\frac{1}{x}\int_0^x\frac{B^2}{8\pi}dx^\prime\right).
\end{equation}
Note that the velocity at any given point is proportional to
the parametre $\lambda$ controlling the weak interaction rate and
to the difference of the {\it average} magnetic pressure to the
left and to the right of this point. This is because, in diffusive
equilibrium, a stronger magnetic pressure corresponds to a lower
density of charged particles, and the magnetic flux can only
spread as the reactions modify the charged-particle density.

Replacing $v_c$ in Eq.~(\ref{bz}), we obtain the
integro-differential equation governing the magnetic field
evolution
\begin{eqnarray}
\label{magn-equati1}
\frac{\partial B}{\partial t}=
-\frac{\partial}{\partial x}\left(v_{c}B\right)=\nonumber\\\frac{\lambda}{dn_{0c}^ 2}\frac{\partial}{\partial x}\left[x(d-x)\left(\frac{1}{d-x}\int_x^d\frac{B^2}{8\pi}dx^\prime-\frac{1}{x}\int_0^x\frac{B^2}{8\pi}dx^\prime\right)B\right].
\end{eqnarray}

The characteristic time-scale from this equation is the same as
estimated in Paper I. If we scale the position variable $x$ in
Eq.~(\ref{magn-equati1}) to the total length of the system $d$,
the time variable  to $t_{ambip}^{weak}$, and the magnetic field
to some characteristic magnetic field (for our subsequent
numerical analysis it is chosen as the maximum of the initial
magnetic field profile), we can write this equation with
dimensionless variables as
\begin{equation}
\label{magn-equati1-nodime} 
\frac{\partial B}{\partial t}=\frac{\partial}{\partial x} \left[
x(1-x)\left(\frac{1}{1-x}\int_x^1B^2dx^\prime
-\frac{1}{x}\int_0^xB^2dx^\prime\right)B\right].
\end{equation}

\subsection{Analytical and numerical solutions}

Below, we solve Eq.~(\ref{magn-equati1-nodime}) by using a
numerical finite-difference scheme. However, and with the goal of
testing the adequate performance of the numerical method, it is
desirable to know some analytical solutions to compare with the
numerical results. Following this philosophy, we try an analytical
solution of Eq.~(\ref{magn-equati1-nodime}) in the form of a
``step'' profile, motivated by the fact that we observed it as a generic asymptotic
state in the evolution of some of the initial profiles that we
study in the following paragraphs. Thus, we make the {\it
Ansatz} 
\begin{equation}
\label{explicit-step1}
B(x,t) = \left\{\begin{array}{lll}
         0 & \text{if} & 0<x < x_1(t),\\
         B_s(t) & \text{if} & x_1(t)<x<1.
         \end{array} \right.
\end{equation}
Using the condition of flux conservation,
$\Phi=B_s(t)(1-x_1(t))=constant$, and replacing
Eq.~(\ref{explicit-step1}) in the dimensionless version of
Eq.~(\ref{ambvelr1}) evaluated at $x=x_1$, we get
\begin{equation}
\label{diffasr1} \frac{d x_1}{dt}=-\Phi^ 2 \frac{x_1}{1-x_1}.
\end{equation}

Integrating Eq.~(\ref{diffasr1}) yields

\begin{equation}
\label{tx1}
t-t_0=-\frac{1}{\Phi^2}\left[\ln\left(\frac{x_1(t)}{x_1(t_0)}\right)
-(x_1(t)-x_1(t_0))\right],
\end{equation}
where $t_0$ is some reference time. Inverting this equation gives $x_1(t)$, while $B_s(t)$ can be
obtained through the condition of flux conservation.

\begin{figure}
\resizebox{8cm}{6cm}{\includegraphics{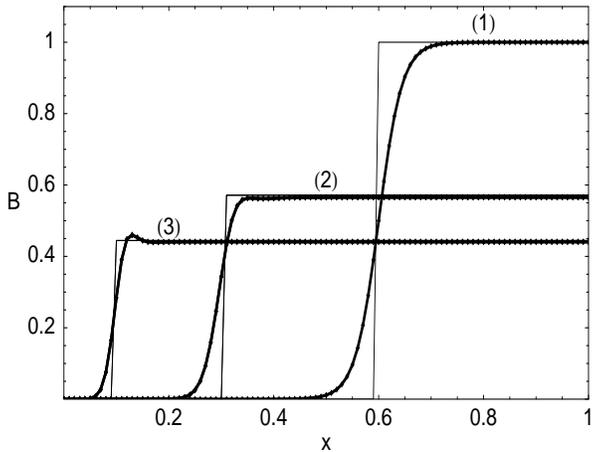}} \caption{The
analytic evolution of an initial magnetic step profile of the form
given by Eq.~(\ref{explicit-step1}) (thin line), and comparison with
the finite-difference evolution of an initial profile given by
Eq.~(\ref{perta}) but using  Eq.~(\ref{magn-equati1-nodime})
(thick line). The time scaling in this figure and in the
following ones in this regime is such that  $t_{ambip}^{(weak)}=1$
and the time progression is labelled as: (1) $t=0$, (2) $t=2.45$,
(3) $t=8.07$. We used the parametres $x_0=0.6$, $t_0=0$, $a=0.04$.
$B_s(0)=1$ and $x_1(0)=0.6$ were chosen so that both profiles
share the same magnetic flux.} \label{g1}
\end{figure}

The step profile Eq.~(\ref{explicit-step1}) presents a sharp
gradient around $x_1(t)$.
Since we are neglecting the ohmic dissipation, any attempt
that we made to use this step profile as an initial condition in
the finite-difference scheme results in a numerical instability.
In order to overcome this difficulty, we compare in Fig.~\ref{g1} the analytical
evolution of Eq.~(\ref{explicit-step1}) with the finite-difference
solution of Eq.~(\ref{magn-equati1-nodime}), using as
initial condition a smooth profile,
\begin{equation}
\label{perta}
B(x,0)=\frac{1}{2}\left[1+\tanh\left(\frac{x-x_0}{a}\right)\right],
\end{equation}
which is a good approximation to Eq.~(\ref{explicit-step1}).

For this comparison, the initial parametres $x_1(t_0=0)$ and $B_s(t=0)$
are chosen so that the step profile given by Eq.~(\ref{explicit-step1})
and the smooth profile given by Eq.~(\ref{perta}) share the same initial magnetic flux.
We see from Fig.~\ref{g1} that, except for the differences around $x_1(t)$,
the evolution of both profiles is similar.

\begin{figure}
\resizebox{8cm}{6cm}{\includegraphics{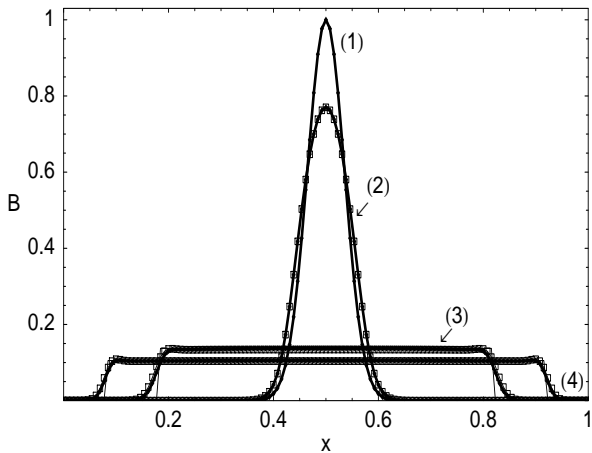}} \caption{The
evolution of a Gaussian initial magnetic field profile given by
Eq.~(\ref{nonlineraBp2}) with $s=20$, and $x_0=0.5$ from the
finite-difference solution of the coupled system of
Eqs.~(\ref{bz}), (\ref{nB}), and (\ref{nc}) (thick line). In
the instant $t=20t_{drag}$ [labelled with (2)] this profile has
evolved to be consistent with the diffusive equilibrium. So, from
this instant we also follow the evolution given by the numerical
finite-difference solution of Eq.~(\ref{magn-equati1-nodime})
(line with empty squares). In addition, from the instant
$t=50t_{ambip}^{(weak)}$ [labelled with (3)], we also compare with
the box-type solution given by Eq.~(\ref{explicit-step}) (thin line),
where $x_1$ at this instant was calculated so that all the
profiles share the same magnetic flux, using the flux conservation
relation $\Phi=B_s(1-2x_1)$. For this simulation, we set as
initial condition for the particle densities $\delta
n_B(x,0)=\delta n_c(x,0)=0$ and the parametres: $n_{0c}/n_{0n}=0.04$, $L/d=0.08$, $\beta=2.0$, and $t_{weak}/t_{drag}\approx 100$. The labels in the figure represent the instants: (1) $t=0$, (2) $t=20t_{drag}=0.1$, (3) $t=50t_{ambip}^{(weak)}=50$, (4) $t=100t_{ambip}^{(weak)}=100$.} 
\label{g2}
\end{figure}

Now, we intend to verify whether the asymptotic evolutionary
equation Eq.~(\ref{magn-equati1-nodime}) correctly describes the
long-term behavior of the magnetic field. This requires a
comparison of the asymptotic evolution of the magnetic field as
given by the full set of Eqs.~(\ref{bz}), (\ref{nB}), and
(\ref{nc}) with the corresponding evolution that arises when using
only Eq.~(\ref{magn-equati1-nodime}). In Fig.~\ref{g2}, we carry
out this comparison for the evolution of a Gaussian initial
magnetic field profile given by
\begin{equation}
\label{nonlineraBp2}
B(x,0)=\exp[-s^2(x-x_0)^2].
\end{equation}

As initial condition for the particle densities we set $\delta
n_B(x,0)=\delta n_c(x,0)=0$. Note that this comparison is valid at
late times, when this profile has evolved to be consistent with
the diffusive equilibrium. So, we start this comparison from the
instant labelled with number $(2)$ in this figure, which
corresponds to $t^*=20t_{drag}$. We see that the asymptotic
behavior of the magnetic field (labels (3) and (4) in the figure)
calculated from these two methods is the same. With the purpose of
characterizing  this asymptotic behavior analytically, we
construct an analytic ``box''-type solution of the form
\begin{equation}
\label{explicit-step}
B(x, t) = \left\{\begin{array}{lll}
         0 & \text{if} & 0<x < x_1(t),\\
         B_s(t) & \text{if} & x_1(t)<x<1-x_1(t), \\
         0 & \text{if} & 1-x_1(t)<x<1.\end{array} \right.
\end{equation}
with
\begin{equation}
\label{diffasr11}
\frac{d x_1}{dt}=-\Phi^ 2 \frac{x_1}{1-2x_1},
\end{equation}
where the magnetic flux is $\Phi=B_s(t)(1-2x_1(t))$, and $t(x_1)$
is given by
\begin{equation}
\label{tx2}
t-t_0=-\frac{1}{\Phi^2}\left[\ln\left(\frac{x_1(t)}{x_1(t_0)}\right)
-2(x_1(t)-x_1(t_0))\right].
\end{equation}
For the instants (3) and (4) in Fig.~\ref{g2}, we show a full line that represents
the solution given by Eq.~(\ref{explicit-step}). We see that this
analytical characterization works well.

\begin{figure}
\resizebox{8cm}{6cm}{\includegraphics{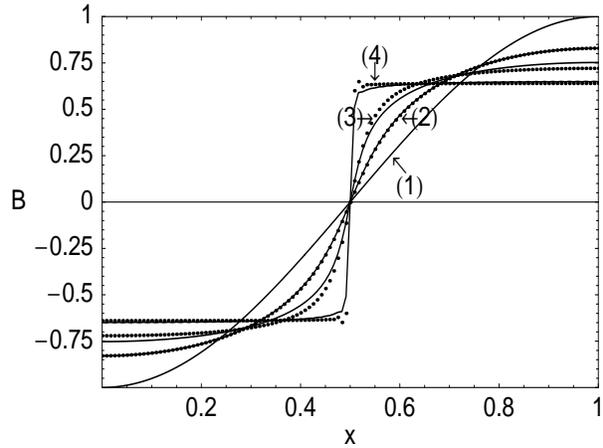}}
\caption{The evolution of Eq.~(\ref{nonlineracosp2}) from the
finite-difference solution of the full coupled system of
Eqs.~(\ref{bz}), (\ref{nB}), and (\ref{nc}) (full line). Similarly
to Fig.~\ref{g2}, in the instant $t=10t_{drag}$ [labelled with
(2)], this profile has evolved to be consistent with the diffusive
equilibrium. So, from this instant we also follow the evolution
given by the numerical finite-difference solution of
Eq.~(\ref{magn-equati1-nodime}) (points). For this simulation we
set again as initial condition $\delta n_B(x,0)=\delta n_c(x,0)=0$
and the same background parametres of Fig.~\ref{g2}. We set
$\beta=1.1$ and $t_{weak}/t_{drag}\approx 100$. The labels in the figure represent (1) $t=0$, (2)
$t=10t_{drag}=0.09072$, (3) $t=t_{ambip}^{(weak)}=1$, (4)
$t=5t_{ambip}^{(weak)}=5$.} 
\label{g3}
\end{figure}

In Fig.~\ref{g3}, we show the evolution of a harmonic initial
magnetic field profile
\begin{equation}
\label{nonlineracosp2}
B(x,0)=-\cos(\pi x).
\end{equation}

In this figure, we compare the evolution of this profile given by
Eq.~(\ref{magn-equati1-nodime}) with that given by the full system
of coupled Eqs.~(\ref{bz}), (\ref{nB}), and (\ref{nc}). For the latter, the initial conditions for the particle
density perturbations are $\delta n_B(x,0)=\delta n_c(x,0)=0$. At
the instant labelled with number (2), $t^*=10t_{drag}$, this
initial profile has evolved to be consistent with the diffusive
equilibrium, therefore, from this instant we also calculate the
evolution of the magnetic field through
Eq.~(\ref{magn-equati1-nodime}). We see in this figure
that the asymptotic behavior of the magnetic field [instant (4)]
calculated from these two methods is the same, which again
verifies the validity of Eq.~(\ref{magn-equati1-nodime}) to model
the asymptotic evolution of the magnetic field.

\begin{figure}
\resizebox{8cm}{6cm}{\includegraphics{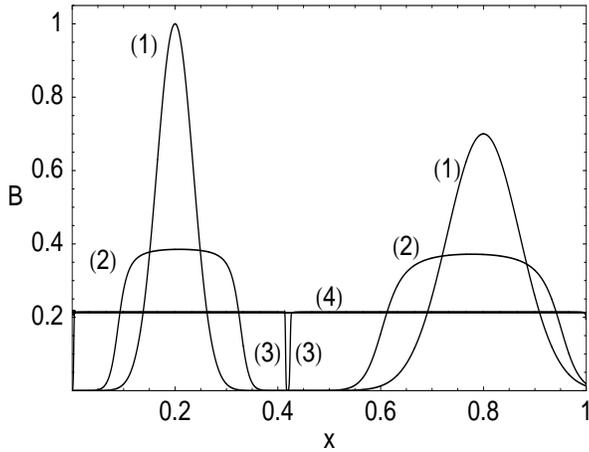}}
\caption{Evolution of a double Gaussian profile of the form
$B(x,0)=\exp[-400(x-0.2)^2]+0.7\exp[-100(x-0.8)^2]$, using
Eq.~(\ref{magn-equati1-nodime}). The labels indicate the time
progression for different instants, (1) $t=0$, (2) $t=4.0$, (3)
$t=60.0$, (4) $t=400.0$.} 
\label{p1arbs1}
\end{figure}

\begin{figure}
\resizebox{8cm}{6cm}{\includegraphics{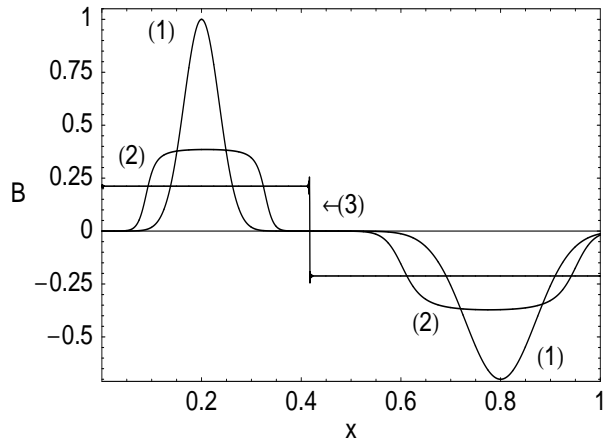}}
\caption{Evolution given by Eq.(\ref{magn-equati1-nodime}) for the
same instants as in Fig.~\ref{p1arbs1}, but for an initial profile
with one of the Gaussians inverted:
$B(x,0)=\exp[-400(x-0.2)^2]-0.7\exp[-100(x-0.8)^2]$}
\label{p1arbs2}
\end{figure}

In summary, the last results confirm the adequacy of
Eq.~(\ref{magn-equati1-nodime}) to describe the asymptotic
evolution of the magnetic field promoted by ambipolar diffusion
but controlled by beta decays. So, in what follows we explore the
evolution of different magnetic profiles using only this equation.
This also has the advantage that we can increase the time-step
with respect to that needed when solving the full set of
Eqs.~(\ref{bz}), (\ref{nB}), and (\ref{nc}) without generating
numerical instabilities. In Figs.~\ref{p1arbs1}-\ref{p1arbs2}, we
show the evolution of different initial profiles. From the
numerical results, we can infer some generic properties:

\begin{itemize}
\item The ambipolar diffusion process tends to spread out the
magnetic flux and the total magnetic flux is conserved as it is
expected from our boundary conditions. \item Contrary to the
normal Ohmic diffusion, in Fig.~\ref{p1arbs2} it is
observed that the magnetic field is not smoothed over all the
space. It becomes uniform only over regions whose boundaries
are points where the magnetic field is zero, and across which it
jumps between values of the same magnitude, but opposite signs,
making the magnetic pressure $B^2/(8\pi)$ equal on both sides.

\item The magnetic field nulls move according to the local values of the charged particle
velocity, as given by Eq.~(\ref{ambvelr1}). This velocity is a
continuous function of $x$, so the flux can be spread out or
compressed, but no reconnection (mutual elimination of opposite
field lines) occurs in the absence of Ohmic dissipation, contrary
to the strong coupling limit to be studied in the next section. Of
course, the formation of steep gradients makes it possible for
Ohmic dissipation and therefore reconnection to occur in a
realistic astrophysical setting.
\end{itemize}

\section{Magnetic field evolution in the strong coupling limit}
\label{asy2}

\subsection{Derivation of the asymptotic evolutionary equation}

In this section, we study the magnetic field evolution in the
opposite limit of the last section, namely, the {\it strong
coupling limit}, $t_{drag} \gg t_{weak}$, in which the
conversion from charged to neutral particles and vice-versa is
essentially instantaneous, but their relative motion is impeded by
a strong mutual collisional drag force. This limit is relevant for
young, hot neutron stars, as well as in the interstellar medium
\citep{Shu-83,BZ-94}.

In order to study the long-term evolution, we consider that the
system has reached both the magneto-hydrostatic and chemical
equilibria. From Eq.~(\ref{vn}), the magneto-hydrostatic
equilibrium condition implies
\begin{equation}
\label{equiltmhd}
n_{0n}\delta \mu_n+n_{0c}\delta \mu_c+\frac{B^2}{8\pi}=h(t),
\end{equation}
where $h(t)$ is an arbitrary function depending only on the time variable. On the other hand, the condition for chemical equilibrium implies
\begin{equation}
\label{chemicalt}
\delta \mu_n-\delta \mu_c=0
\end{equation}
Combining Eqs.~(\ref{equiltmhd}) and (\ref{chemicalt}), we obtain
\begin{equation}
\label{dncr2} \delta \mu_c=\delta
\mu_n=\frac{1}{n_{0B}}\left[h(t)-\frac{B^2}{8\pi}\right].
\end{equation}
As done in Eq.~(\ref{orderan}), we can compare the time derivative
to one of the spatial derivatives in Eq.~(\ref{nB}),

\begin{equation}
\label{orderan2} 
\left|\partial\delta n_B\over\partial
t\right|\sim \frac{\delta n_B}{t_{ambip}^{(drag)}} \ll
\frac{n_{0c}v_c}{L}\sim n_{0c}\left|\partial v_c\over\partial
x\right|.
\end{equation}
Thus, we can neglect the time derivative of $\delta n_B$  in
comparison with the spatial derivative term in Eq.~(\ref{nB}) and
write
\begin{equation}
\label{vnvx2}
n_{0n}\frac{\partial v_n}{\partial x}=-n_{0c}\frac{\partial v_c}{\partial x}.
\end{equation}

Integrating Eq.~(\ref{vnvx2}) between $x=0$ and an inner point $x$
with the boundary conditions $v_c(0,t)=v_n(0,t)=0$, and using
Eq.~(\ref{dncr2}), we obtain
\begin{equation}
\label{vAsegr}
v_c=-\frac{n_{0n}}{n_{0c}n_{0B}^{2}\gamma_{cn}}\frac{\partial}{\partial x}\left(\frac{B^2}{8\pi}\right).
\end{equation}
We see from Eq.~(\ref{vAsegr}) that the Lorentz force drives the motion of magnetic flux and charged particles at the velocity $v_c$, which is controlled by the inter-particle collisions through the factor $1/\gamma_{cn}$. After including Eq.~(\ref{vAsegr}) in  Eq.~(\ref{bz}), we obtain the equation for the magnetic field evolution as
\begin{equation}
\label{magn-equati2}
\frac{\partial B}{\partial t}=-\frac{\partial}{\partial x}\left(v_{c}B\right)=\frac{n_{0n}}{8\pi n_{0c}n_{0B}^{2}\gamma_{cn}}\frac{\partial}{\partial x}\left[B\frac{\partial B^2}{\partial x}\right]
\end{equation}

Scaling the time variable to $t_{ambip}^{(drag)}$ and the position variable to the total length of the system $d$, this equation can be written in dimensionless form as
\begin{equation}
\label{magn-equatiadimensi2}
\frac{\partial B}{\partial t}=\frac{\partial}{\partial x}\left[B\frac{\partial B^2}{\partial x}\right].
\end{equation}

Eq.~(\ref{magn-equatiadimensi2}) belongs to a group of {\it
non-linear diffusion equations} or {\it porous medium equations}
whose mathematical properties have been studied by different
authors (see e.g. \citealt{T-76,V-07}).  \citet{Shu-83}
obtained it for ambipolar diffusion in interstellar gas, in a
similar, strong-coupling regime as considered here, but ignoring
the pressure of the charged particles, while allowing for
arbitrarily large density perturbations, and using a Lagrangian
coordinate system moving with the neutral fluid. It was re-derived
under slightly different assumptions by \citet{BZ-94}.

\subsection{Analytical and numerical solutions}
\subsubsection{Exact analytical solution}
As an analytical solution of Eq.~(\ref{magn-equatiadimensi2}) that will allow us to validate our numerical results, we follow \citet{T-76} and consider
\begin{equation}
B(x, t) = \left\{\begin{array}{lll}
         0 & \text{if} & |x-x_0| > a_s(t)\\
         B_s(t) \sqrt{1-\left(\frac{x-x_0}{\overline a_s(t)}\right)^2} &  \text{if} & |x-x_0| \leq a_s(t).\\
         \end{array} \right. 
 \label{expicit-nonlin-diff}         
\end{equation}
This solution is a semi-ellipse in the $(x,B)$ plane, centered at $(x_0,0)$, with semi-axes $a_s(t)$ and $B_s(t)$. Replacing Eq.~(\ref{expicit-nonlin-diff}) in
Eq.~(\ref{magn-equatiadimensi2}) and using the flux conservation
condition $\Phi=(\pi/2)B_s(t)a_s(t)=\mathrm{constant}$, we obtain
the differential equation  for $a_s(t)$, namely,
\begin{equation}
\label{diffas} \frac{da_s}{dt}=\frac{8\Phi^ 2}{\pi^2 a_s^3},
\end{equation}
with the solution
\begin{equation}
\label{astt}
a_s(t)=a_s(t^*)\left[1+\frac{8B_s^2(t^*)}{a_s^2(t^*)}(t-t^*)\right]^{1/4},
\end{equation}
where $t^*$ is a reference time.

\begin{figure}
\resizebox{8cm}{6cm}{\includegraphics{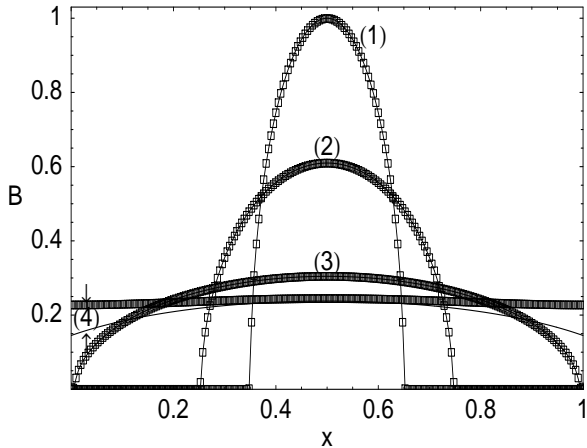}}
\caption{Evolution of an initial ``elliptic'' profile of the form
Eq.~(\ref{expicit-nonlin-diff}) from
Eqs.~(\ref{diffas}),(\ref{astt}) (line), and evolution given by
the numerical finite-difference solution of
Eq.~(\ref{magn-equatiadimensi2}) (empty squares). The time
instants are (1) $t=0$, (2) $t=0.02$, (3) $t=0.3$, (4) $t=0.9$}
\label{g5}
\end{figure}

In Fig.~\ref{g5}, we compare this analytical solution to the corresponding evolution obtained from the finite-difference
numerical integration of Eq.~(\ref{magn-equatiadimensi2}). 

We see
that there is a good agreement between the two methods except at
the last instant, labelled with $(4)$ in the figure. This is
expected, since at this instant the elliptic part of the profile
has crossed the boundaries of the system, so the analytical
solution is no longer valid. Note also that the latter does not conserve magnetic flux once the ends of the
ellipse have reached the boundaries. We also see at the instant
labelled with $(4)$ how the magnetic field given by
Eq.~(\ref{magn-equatiadimensi2}) is almost homogeneously
distributed across the system, as expected from the magnetic-flux
conserving boundary conditions that we are using for this
equation. Thus, the finite-difference method that we are using to
solve Eq.~(\ref{magn-equatiadimensi2}) appears to be working well.
\subsubsection{Numerical explorations}
Hereafter, and with the main goal of establishing a comparison, we
follow the evolution of the same initial profiles that
we used in the previous section, where we dealt with the opposite
regime.

\begin{figure}
\resizebox{8cm}{6cm}{\includegraphics{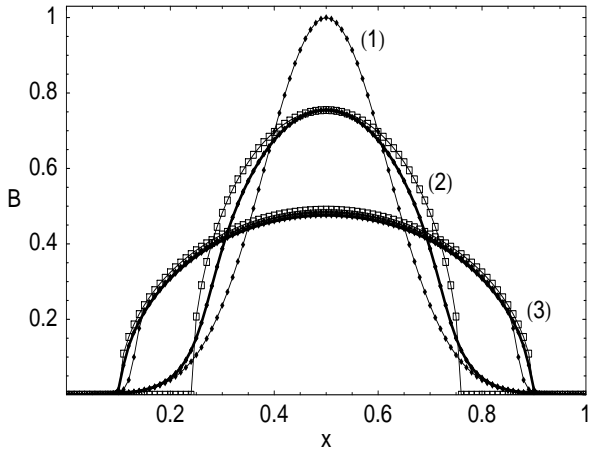}}
\caption{Evolution of a Gaussian initial field profile of the form
given by Eq.~(\ref{nonlineraBp2}) with $s=6$, and $x_0=0.5$, from
the finite-difference solution of the coupled system of
Eqs.~(\ref{bz}), (\ref{nB}), and (\ref{nc}) (line with points). For
this simulation we set as intitial condition for the particle
densities $\delta n_B(x,0)=\delta n_c(x,0)=0$. In the instant
$t=10t_{weak}$ [labeled with (2)], this profile has evolved to be
consistent with the chemical equilibrium. So, from this instant we
also follow the evolution given by the numerical
finite-differences solution of Eq.~(\ref{magn-equatiadimensi2})
(thick line) and the analytical solution (line with empty squares)
Eq.~(\ref{expicit-nonlin-diff}) whose parametres were chosen in a
such way as to share the same magnetic flux. We used the
parametres $\beta=1.1$, $t_{drag}/t_{weak}\approx 100$. 
The labels in the figure represent the instants: (1) $t=0$,
(2) $t=10t_{weak}$, (3) $t=t_{ambip}^{(drag)}=1$. We used the
same background parametres $n_{0c}/n_{0n}=0.04$.} 
\label{g6}
\end{figure}

In Fig.~\ref{g6}, we show the evolution of an initial Gaussian
magnetic field profile given by Eq.~(\ref{nonlineraBp2}), as given
by the full coupled system of Eqs.~(\ref{bz}), (\ref{nB}), and
(\ref{nc}) with $\delta n_B(x,0)=\delta n_c(x,0)=0$. At the
instant (2), $t^*=10t_{weak}$, the system should be very close to chemical equilibrium.
Thus, from this instant onwards, Eq.~(\ref{magn-equatiadimensi2})
should be valid, so we also follow the evolution of the magnetic
field obtained from the finite-difference solution of this
equation. We see that the asymptotic behavior of the magnetic
field (labels (2) and (3) in the figure) calculated from these two
models is the same. In addition, for the instants 
(2) and (3), we show for comparison the analytical solution
[Eq.~(\ref{expicit-nonlin-diff})].

We see that this explicit solution is an adequate description of
the asymptotic evolution.

\begin{figure}
\resizebox{8cm}{6cm}{\includegraphics{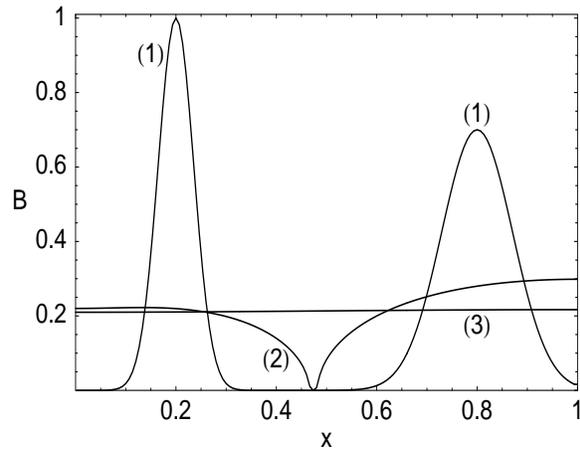}}
\caption{Evolution according to Eq.~(\ref{magn-equatiadimensi2})
of an initial field
$B(x,0)=\exp(-400(x-0.2)^2)+0.7\exp(-100(x-0.8)^2)$. The label
numbers show the time progression for different instants, (1)
$t=0$, (2) $t=0.3$, (3) $t=3.0$} \label{g66}
\end{figure}

\begin{figure}
\resizebox{8cm}{6cm}{\includegraphics{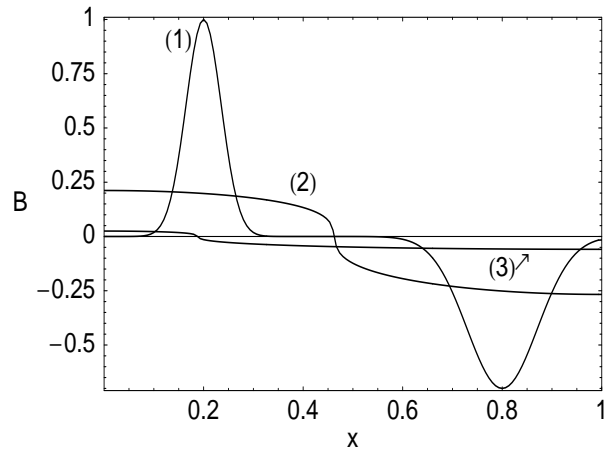}}
\caption{Evolution according to Eq.~(\ref{magn-equatiadimensi2})
of an initial field
$B(x,0)=\exp(-400(x-0.2)^2)-0.7\exp(-100(x-0.8)^2)$.The label
numbers show the time progression for different instants, (1)
$t=0$, (2) $t=0.5$, (3) $t=45.0$} \label{g666}
\end{figure}

The previous results confirm the adequacy of
Eq.~(\ref{magn-equatiadimensi2}) in describing the asymptotic
evolution of the magnetic field promoted by ambipolar diffusion
and controlled by inter-particle collisions. We now examine the
evolution of different magnetic profiles using only this equation.
In Figs.~\ref{g66}-\ref{g70} we see the evolution of different
magnetic field profiles. From these numerical results, we can
infer the following properties:
\begin{itemize}
\item As in the weak coupling limit, the ambipolar diffusion process acts in
a tendency to spread out the magnetic flux. Again, as expected,
the total magnetic flux is conserved. \item At the null
points, the magnetic field vanishes continuously, but with
a high (possibly infinite) derivative, as expected from the
example of the solution given by Eq.~(\ref{expicit-nonlin-diff}).

\item Contrary to the weak coupling limit, in Fig.~\ref{g666}, it is observed that the magnetic flux is
not preserved in each one of the regions separated by the magnetic
null points. Therefore, there is a transfer of magnetic flux
through these null points, leading to reconnection of magnetic
field lines even in the absence of Ohmic diffusion, as will be
discussed further below.
\end{itemize}

\subsubsection{Fixed singularity}
In order to understand the formation and behaviour of the
singularities at the null points, we first consider a magnetic
field with a finite derivative at a null point $x=x_0$, and follow
the evolution of its derivative $p(t)$ using the {\it Ansatz}
\begin{equation}
\label{pend} B(x,t)=p(t)(x-x_0).
\end{equation}
The transport velocity,
\begin{equation}
v_c=-\frac{\partial}{\partial x}\left(B^2\right)=-2p^2(x-x_0),
\end{equation}
vanishes at the null point as long as $p$ is finite, so there is
no reconnection at the null point in this regime. From Eq.~(\ref{magn-equatiadimensi2}), the differential equation for $p(t)$ is $dp/dt=4p^3,$ with the solution$p(t)=p(0)/\sqrt{1-8p^2(0)t},$ which diverges in a finite time $t_\infty=1/(8p^2(0))$. \citet{BZ-94} found that this divergence (which they identified
numerically) can lead to a stationary solution for the magnetic
field. We re-derive it by noting that, to have $\partial
B/\partial t=0$ in Eq.~(\ref{magn-equatiadimensi2}), the term in
parenthesis on the right-hand side, which corresponds to (minus)
the ``flux of flux'' (amount of magnetic flux crossing any point
$x$ per unit time), must be uniform in space,
\begin{equation}
\label{standing}
B\frac{\partial B^2}{\partial x}=-v_cB\equiv -F=\mathrm{constant},
\end{equation}
leading to
\begin{equation}
\label{asyu} 
B(x)=\left[-\frac{3}{2}F(x-x_0)\right]^{1/3}.
\end{equation}
The infinite derivative at $x=x_0$ allows $F$ to remain finite at
this point, so there is magnetic flux crossing the singularity and
causing reconnection, without having included Ohmic diffusion in
the model. Of course, this simple solution is not compatible with
our boundary conditions, which were constructed so as to enforce
$F=0$ at the boundaries. Therefore, for our numerical solutions,
the reconnection must produce a decrease in the absolute value of
the magnetic flux in each of the regions separated by the
singularity, and therefore in the amplitude of the magnetic field. 

In order to explore this behaviour, we will study the evolution of an initial magnetic field profile of the form
\begin{equation}
\label{sep0}
B(x,0)=\cos(\pi x).
\end{equation}
This magnetic field profile has a null at $x=1/2$ and is an odd function respect to this point, a property that is preserved by the evolution according to Eq.~(\ref{magn-equatiadimensi2}). Thus, in particular, the position of the null point will remain fixed. Based on this fact, we try a solution of Eq.~(\ref{magn-equatiadimensi2}) by separation of variables in the form
\begin{equation}
\label{sep1}
B(x,t)=f(t)g(x),
\end{equation}
with $f(t)$ and $g(x)$ satisfying
\begin{equation}
\label{sep2}
\frac{df}{dt}=-Cf^{3},
\end{equation}
and
\begin{equation}
\label{sep3}
\frac{d}{dx}\left[g^2 \frac{dg}{dx}\right]=-\frac{C}{2}g,
\end{equation}
where $C$ is a separation constant. The magnetic flux between the left boundary of the system and the null point at $x=1/2$ is 
\begin{equation}
\label{sep5}
\Phi^+(t)=\int_{0}^{1/2}B(x,t)dx.
\end{equation}
By symmetry, the flux in the other half of the interval will be
\begin{equation}
\Phi^-(t)=\int_{1/2}^{1}B(x,t)dx=-\Phi^+(t).
\end{equation}
If we normalize,
\begin{equation}
\label{sep67}
\int_{0}^{1/2}g(x)dx=1, 
\end{equation}
we will have $f(t)=\Phi^+(t)$. 

In order to solve Eq.~(\ref{sep3}) subject to the boundary conditions of Eq.~(\ref{boundaries}), we use the new variable $u(x)=g^3(x)$, which must satisfy the conditions $(du/dx)_{x=0}=(du/dx)_{x=1}=0$ and $u(1/2)=0$. Thus, Eq.~(\ref{sep3}) is rewritten as
\begin{equation}
\label{sepu1}
\frac{d^2u}{dx^2}=-\frac{3}{2}Cu^{1/3}.
\end{equation}
Combining Eq.~(\ref{sepu1}) with Eq.~(\ref{sep67}), we obtain the auxiliary condition $(du/dx)_{x=1/2}=-3C/2$, which indicates that the magnetic field close to the null point has the shape given by Eq.~(\ref{asyu}). The numerical solution of Eq.~(\ref{sepu1}) must satisfy conditions at different points, which can be intricate from the numerical point of view. Thus, we rescale of our variables as $u=Uv$ and $x=Xy$, where $U$ and $X$ are constants to be determined. This allows us to write Eq.~(\ref{sepu1}) as
\begin{equation}
\label{sepu2}
\frac{d^2v}{dy^2}=-v^{1/3},
\end{equation}
which requires $C=(2/3)X^{-2}U^{2/3}$. The boundary conditions can be set as $v(y=0)=1$, $(dv/dy)_{y=0}=0$. From the numerical solution of Eq.~(\ref{sepu2}) we obtain $v(y_0)=0$ at $y_0=1.46$. Since the null point of $u$ is at $x_0=1/2$, we get $X=x_{0}/y{_0}=0.34$. Using $u=Uv$ and the conditions on the first derivatives, $(du/dx)_{x=1/2}=-3C/2$ and $(dv/dy)_{y=y_0}\equiv p_0=-1.20$ (obtained numerically), we get $U=-1/(Xp_0)^3=14.0$. Finally, comparing Eqs.~(\ref{sepu1}) and (\ref{sepu2}) we obtain $C=(32/3)(y_0^4/p^2)=33.7$, which determines the full evolution of the magnetic flux in each half of the interval, which is given according to Eq.~(\ref{sep2}) as   
\begin{equation}
\label{sep4}
\Phi^+(t)=\frac{\Phi^+(0)}{\sqrt{1+2 C \left[\Phi^+(0)\right]^2 t}}~\underset{t \to \infty}{\to} \frac{1}{\sqrt{2Ct}}.
\end{equation}
The flux calculated from the numerical solution of Eq.~(\ref{magn-equatiadimensi2}) agrees with Eq.~(\ref{sep4}) with a percentage error of order $0.4\%$. Eq.~(\ref{sep4})  implies that positive magnetic flux coming from the left of the singularity annhilates with the negative flux coming from the right. This is reconnection in the absence of Ohmic resistivity.

\begin{figure}
\resizebox{8cm}{6cm}{\includegraphics{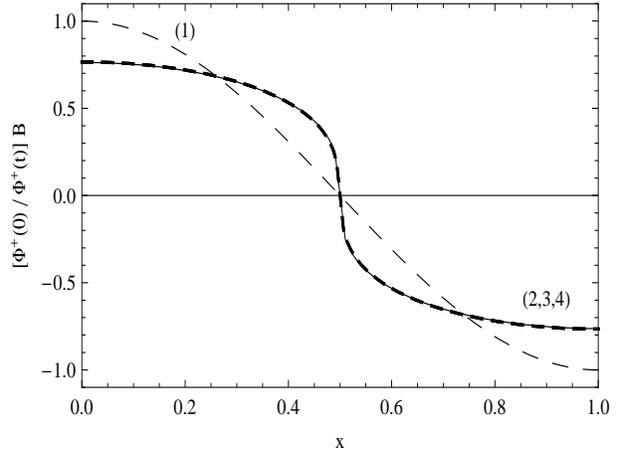}}
\caption{Evolution of an initial magnetic field $B(x,0)=\cos(\pi x)$. The label numbers show the time progression for different instants, (1) $t=0$, (2) $t=0.5$, (3) $t=5$, (4) $t=50$. The dashed lines represent the evolution as given by Eq.~(\ref{magn-equatiadimensi2}) while the full line is the asymptotic solution given by Eq.~(\ref{sep1}). It follows that the solution of Eq.~(\ref{magn-equatiadimensi2}) converges after an initial transient to Eq.~(\ref{sep1}). $\Phi^+(t)$ is the positive magnetic on the left of the singularity $x=0.5$ [see Eq.~(\ref{sep4})] and $\Phi^+(0)=1/\pi$.}
\label{g70}
\end{figure}

In Fig.~\ref{g70} we observe that the magnetic field as given by the numerical solution of Eq.~(\ref{magn-equatiadimensi2}) converges after a initial transient to the field of Eq.~(\ref{sep1}) (obtained from the solution of Eq.~(\ref{sepu1}). 

This magnetic field profile has a strong divergence of the velocity field at the null point, and we are assuming that the excess charged particles instantaneously recombine when they reach this point, which is not realistic even in the ``strong coupling limit'', in which $t_{weak} \ll t_{drag}$. In other words, even if the beta decays can be considered as instantaneous everywhere else in the system, this approximation will break down close enough to the singularity. The behaviour close to the singularity, including both a fast, but finite recombination rate and a small, but finite Ohmic diffusivity, was discussed in detail by \citet{H-03a,H-03b}.

\section{Conclusions}
\label{conc}

We have studied the asymptotic magnetic field evolution promoted
by ambipolar diffusion in a one-dimensional geometry 
for two opposite, limiting regimes. In the {\it weak coupling
limit}, in which neutral and charged particles drift easily
with respect to each other, the bottleneck for the evolution is
the conversion of one species into another, which is required in
order to eliminate the charged-particle pressure gradients caused
by the magnetic field, which impede its evolution. In the {\it
strong coupling limit}, conversions are easy, but the
inter-particle collisions are the corresponding bottleneck. 
In molecular clouds, the second regime appears to be generally
relevant \citep{Shu-83,BZ-94}. Neutron stars in their hot, early
phase will also be in the strong-coupling regime, and evolve to
the weak coupling regime as they cool.

In the {\it weak coupling limit}, the magnetic field evolution is
described by a non-linear, partial integro-differential equation,
while in the {\it strong coupling limit} this evolution is
described by a non-linear diffusion equation. We made numerical
simulations of the evolution of different initial magnetic
field profiles in each of these limits and found agreement between
our numerical results and some analytic solutions that can be
found for these differential equations.

From our results, we infer that, in both limits, the ambipolar
diffusion process operates in a tendency to spread out the
magnetic flux, but contrary to the normal Ohmic diffusion this
process asymptotically produces singular points with sharp
magnetic field gradients. These sharp gradients develop around
those points where the magnetic field is null, and separate
regions of magnetic fields with opposite signs. We observe some
generic properties of this process, as follows:

In the {\it weak coupling limit}, the resulting
discontinuities can be modelled as step solutions
[Eqs.~(\ref{explicit-step}), (\ref{explicit-step1})]. The
asymptotic magnetic field is spatially uniform in each of the
regions separated by these singularities, and its absolute value
(and thus the magnetic pressure) is the same in each region.
Ambipolar diffusion by itself does not cause magnetic flux
transfer (and thus reconnection) across the singularities, but the
associated current sheets will easily be dissipated by Ohmic
diffusion, so reconnection will occur in a realistic system.
In the {\it strong coupling limit}, at the singular points
the magnetic field vanishes continuously but with infinite spatial
derivative. Ambipolar diffusion acts in a tendency to spread out the magnetic
flux, but, contrary to the weak coupling limit, the magnetic flux
is not preserved in each one of the regions separated by the
magnetic null points. Therefore, there is a transfer of magnetic
flux through these null points, i.~e. reconnection without Ohmic
resistivity (but see \citealp{H-03a,H-03b}).

The main limitation in applying the present formalism to realistic
systems (either neutron stars or molecular cloud cores) is the
very restrictive, one-dimensional geometry. An extension to more
realistic geometries (i.~e., axial symmetry) will be attempted in
further work.

\section*{Acknowledgments}
We are grateful to Mar\'ia Cristina Depassier and Crist\'obal
Petrovich for valuable information about the non-linear diffusion
equation. We also thank the referee for a very helpful report. This work was financed by the Gemini-CONICYT Fund,
project $N^0$ 32070014; FONDECYT regular projects 1060644 and
1070854, the FONDAP Center for Astrophysics (15010003), Proyecto
Basal PFB-06/2007 and the joint project  ``Estudio Computacional del decaimiento de campos magn\'eticos en estrellas de neutrones'' between Universidad de
Medell\'in (Summa Group), Pontificia Universidad Cat\'olica de
Chile and Universidad de Chile. The postdoctoral stay of J.H.H. at
Astrophysikalisches Institut Potsdam was possible due to the
financial support of Deutscher Akademischer Austauschdienst (DAAD)
- Germany and Comisi\'on Nacional de Investigaci\'on Cientifica y
Tecnol\'ogica (CONICYT) - Chile through the postdoctoral
fellowship $N^0$ A0772255-2007-07-DOCDAAD-25. We also thank the FONDECYT International Cooperation Project 7090020.


\begin{thebibliography}{99}
\bibitem[\protect\citeauthoryear{Aguilera, Pons \& Miralles}{2008}]{APM-08} Aguilera D.N., Pons J.A. \&  Miralles J.A. 2008, A\&A, 486, 255
\bibitem[\protect\citeauthoryear{Akmal, Pandharipande \&  Ravenhall}{1998}]{APR-98} Akmal A., Pandharipande V.R. \&  Ravenhall D.G. 1998, Phys. Rev. C, 58, 1804
\bibitem[\protect\citeauthoryear{Arras, Cumming \& Thompson}{2004}]{ACT-04} Arras P., Cumming A., \& Thompson C., 2004, ApJ, 608, L49
\bibitem[\protect\citeauthoryear{Baym, Pethick, \& Pines}{1969}]{Bay-69} Baym G., Pethick C., \& Pines D. 1969, Nat, 224, 872
\bibitem[\protect\citeauthoryear{Brandenburg \& Zweibel}{1994}]{BZ-94} Brandenburg A., Zweibel E.G, 1994, A\&A, 427, 91
\bibitem[\protect\citeauthoryear{Brandenburg \& Zweibel}{1995}]{BZ-95} Brandenburg A., Zweibel E.G, 1995, ApJ, 427, 91
\bibitem[\protect\citeauthoryear{Duncan \& Thompson}{1992}]{DT-92} Duncan R.C., Thompson C., 1992, ApJ, 392, L9
\bibitem[\protect\citeauthoryear{Galli \& Shu}{1993}]{GS-93} Galli D., \& Shu F.H., 1993, ApJ, 417,220
\bibitem[\protect\citeauthoryear{Goldreich \& Reisenegger}{1992}]{GR-92} Goldreich P., \& Reisenegger A. 1992, ApJ, 395, 250
\bibitem[\protect\citeauthoryear{Haensel, Urpin, \& Yakovlev}{1990}]{HUY-90} Haensel P., Urpin V.A., \& Yakovlev D.G. 1990, A\&A, 229, 133
\bibitem[\protect\citeauthoryear{Harrison}{1991}]{H-91} Harrison E. 1991, MNRAS, 248, 419
\bibitem[\protect\citeauthoryear{Heitsch \& Zweibel}{2003a}]{H-03a} Heitsch F. \& Zweibel E.G. 2003a, ApJ, 583, 229
\bibitem[\protect\citeauthoryear{Heitsch \& Zweibel}{2003b}]{H-03b} Heitsch F. \& Zweibel E.G. 2003b, ApJ, 590, 291
\bibitem[\protect\citeauthoryear{Hoyos, Reisenegger, \& Valdivia}{2008}]{HRV-08} Hoyos J., Reisenegger A., \& Valdivia J.A. 2008, A\&A, 487, 789
\bibitem[\protect\citeauthoryear{Jones}{1987}]{J-87} Jones P.B. 1987, MNRAS, 228, 513
\bibitem[\protect\citeauthoryear{Mestel}{1966}]{M-66} Mestel L. 1966, MNRAS, 133, 265
\bibitem[\protect\citeauthoryear{Mestel \& Spitzer}{1956}]{MestelSpitzer56} Mestel L. \& Spitzer L. 1956, MNRAS, 116, 503 
\bibitem[\protect\citeauthoryear{Mouschovias}{1991}]{M-91} Mouschovias T., 1991, ApJ, 371, 296
\bibitem[\protect\citeauthoryear{Mouschovias \& Paleologou}{1981}]{MouschoviasPaleologou81} Mouschovias T. Ch. \& Paleologou E. V. 1981, ApJ, 246, 48
\bibitem[\protect\citeauthoryear{Niebergal, Ouyed, \& Leahy}{2006}]{NOL-06} Niebergal B., Ouyed R., \& Leahy D. 2006, ApJ, 646, L17
\bibitem[\protect\citeauthoryear{Pethick}{1992}]{P-92} Pethick C.J., 1992, in Structure and Evolution of Neutron Stars, ed. D.Pines, R. Tamagaki,\& S.Tsuruta (Redwood City: Addison-Wesley), 115
\bibitem[\protect\citeauthoryear{Reisenegger}{2009}]{R-09} Reisenegger A., 2009, A\&A, 449, 557
\bibitem[\protect\citeauthoryear{Shu}{1983}]{Shu-83} Shu F. H. 1983, ApJ, 273, 202
\bibitem[\protect\citeauthoryear{Spitzer}{1978}]{S-78} Spitzer L. 1978, Physical Processes in the Interstellar Medium, Willey, New York
\bibitem[\protect\citeauthoryear{Thompson \& Duncan}{1996}]{TD-96} Thompson C., Duncan R.C., 1996, ApJ, 473, 322
\bibitem[\protect\citeauthoryear{Tuck}{1976}]{T-76} Tuck B. 1976, J. Phys. D: Appl. Phys., 9, 1559
\bibitem[\protect\citeauthoryear{Vasquez}{2007}]{V-07} V\'azquez J.L., 2007, The Porous Medium Equation: Mathematical Theory, Oxford Mathematical Monographs, Oxford Science Publications, Oxford University Press Inc., New York
\bibitem[\protect\citeauthoryear{Zweibel}{1989}]{Z-89} Zweibel E.G. 1989, ApJ, 340, 550
\end{thebibliography}
\end{document}